# Modeling of polymer phase transition from crystalline to conformationally disordered phase

*Vadim V. Atrazhev, Dmitry V. Dmitriev[1], and Vadim I. Sultanov*

N. M. Emanuel Institute of Biochemical Physics of the Russian Academy of Sciences, 4 Kosygin str., Moscow, 119334, Russia

ABSTRACT: A physics-based analytical model describing the phase transition from crystalline to conformationally disordered (condis) crystalline phase is developed. In the model, the free energy is written as a function of temperature and the lattice parameter (mean distance between neighboring chains). It consists of two contributions: elastic and conformational. The elastic contribution describes the interaction between neighboring chains, while the conformational part takes into account the conformation of one chain inside the potential tube, formed by the neighboring chains. To verify this approach, polyethylene—the simplest polymer possessing the condis phase—was chosen as a modeling object. Previous experiments and molecular dynamics simulations show that the typical conformation of a polymer chain in a crystalline phase consists mainly of *trans* dihedrals and a small fraction of *gauche* dihedrals, which can be considered as defects of the crystalline lattice. These defects displace the chain inside the tube thus increasing the potential energy. The energy required to form such a defect decreases rapidly with increasing distance between neighboring chains. This leads to a first-order phase transition at a certain temperature to the condis phase, in which distance between neighboring chains is large and a fraction of *gauche* dihedrals is high. This physical picture of the phase transition is described by the proposed analytical model, the parameters of which were calibrated against the results of molecular dynamics simulations for atmospheric pressure. The model predictions for the pressure of 500 atm and 1000 atm are in perfect agreement with the results of molecular dynamics simulations.

---

[1] Contact author: dmitriev@deom.chph.ras.ru

1. INTRODUCTION

Conformationally disordered (abbreviated condis) crystals are a type of mesophases, i.e. states of matter that combine some properties of a crystal with some properties of a liquid. According to the Wunderlich classification [1], there are three types of mesophases (besides their glassy analogs): liquid crystals, plastic crystals, and condis crystals. Liquid crystals combine the ordered orientation of molecules with disorder in the location of their centers of mass. Plastic crystals are the opposite of liquid crystals: the centers of mass of the molecules are located at the nodes of the crystal lattice, whereas the orientation of the molecules is disordered.

If the molecule is not rigid, the arrangement of its atoms in space is determined not only by the location of its center of mass and its orientation as a whole, but also by the mutual orientation of its parts, i.e. conformation. Disorder in the mutual orientation of parts can also combine with order in other aspects. In the case of polymer condis crystals, the positional and orientational orders are present only partially, in some specific sense. Every polymer chain is confined to a cylindrical neighborhood of some axis. The orientational order is that these axes are parallel for all chains— while the transverse orientation is chaotic along a chain and not consistent between different chains. The positional order is that these axes form a hexagonal close-packed arrangement resulting in a characteristic peak in X-ray diffraction patterns—despite the fact that the atoms do not reside at the nodes of any crystal lattice.

Unfortunately, Wunderlich's terminology is still not as widely accepted as it deserves. Condis phases are often called in the literature hexagonal [2–4], pseudohexagonal [5,6], or columnar [7,8] (due to some analogy with the discotic columnar liquid crystals). Moreover, since the condis phase is not typically included in university curriculums and thus not widely known in scientific community, some researchers, when faced with a condis phase, invent their own term for it or describe it in words without giving any name (see e.g. [9]).

The polymers in which condis phases were found are quite diverse [10–22]. These are carbon-chain, heterochain, and noncarbon-chain polymers (including in the backbone atoms of such elements as O, N, P, Si, Ge, and B). The backbone also can contain double bonds, carbonyl, carboxyl, or amide carbon, and such rigid constructs as benzene and thiophene rings and norbornane and m-carborane skeletons. As new research emerges, the number of known polymer condis crystals will grow. Such a diversity of backbones and thus of sets of conformations that they can take along with different polarity, polarizability and ability to form hydrogen bonds means that the unified physicochemical description of all polymer condis crystals is hardly possible. A physics-based model presented in this paper is developed for polyethylene and cannot be immediately used for other polymers. However, we believe that the developed approach can be extended to other polymers, for example, to ferroelectric copolymers of vinylidene difluoride.

Polyethylene (or polymethylene) is the simplest polymer known. The condis phase of polyethylene was first discovered by Bassett and Turner [23] at elevated pressure (> 400 MPa) for samples with molecular weight $M_w$ = 1–2×10$^5$. Later, Pennings and Zwijnenburg [24] produced the fibers from a dilute solution of much longer polyethylene chains ($M_w \approx 1.5 \cdot 10^6$) with help of Couette-flow. These fibers contained extended-chain crystals with macroscopic length and produced a condis phase at atmospheric pressure around 425 K. The entropies of phase transition at elevated and normal pressure somewhat differ and amount to 7.1 and 6.4 J K$^{-1}$ (mol CH$_2$)$^{-1}$ respectively. [10]

Polyethylene crystals (all three known modifications) consist of planar chains in all-*trans* conformation, since hydrogen atoms are small and do not make steric hindrance for this. Even moderate (< 10°) deviations from 180° dihedral angles would be enough to produce a condis phase with chains consisting of dynamically changing helical segments of different chirality (as happens in polytetrafluoroethylene [25]). However, it is known that essential for condis polyethylene are *gauche* backbone dihedrals. Polyethylene crystals contain some insignificant number of *gauche* dihedrals, which are conceived as defects of crystalline lattice. Tashiro et al. [26] on the basis of their Raman spectroscopic measurements estimate the fraction of *gauche* dihedrals in elevated pressure condis polyethylene as "higher than 20%". While the fraction of *gauche* dihedrals in polyethylene melt was estimated as 40% [27]. In our previous molecular dynamics (MD) study [28], at the temperature 25 K below the phase transition, the fraction of *gauche* dihedrals in crystalline polyethylene was 0.2%, just before the transition 1.2%, in condis crystal just after the transition 19%, just before melting 33%, and 41% in the melt just after melting. The consequence of the emergence of large quantity of *gauche* dihedrals is shrinking of the crystal in the longitudinal direction on transition to condis phase, while two transversal dimensions grow.

Our previous work focused on the modeling of the condis phase in ferroelectric copolymers based on vinylidene difluoride (VDF). It was proposed and confirmed by MD simulations [29] that the phase transition between the ferroelectric β-phase and the paraelectric condis phase is responsible for the large electrocaloric effect in such copolymers. The electrocaloric effect is an adiabatic temperature change of a dielectric material when an external electric field is applied or removed. The presence of permanent dipole moment in the monomers of VDF-based copolymers allows to govern the phase transition temperature in such copolymers by an external electric field. The electrocaloric effect (the value of adiabatic temperature change) for perfect crystals of three polymers was calculated by MD simulations in [29]. For quantitative modeling of the electrocaloric effect for realistic amorphous-crystalline polymers, an analytical model based on Landau's free energy functional was proposed in [30] and was further developed in [31]. However, Landau's approach is phenomenological, and the coefficients of the free energy functional in this approach do not have a specific physical meaning and cannot be related to the structure and composition of the polymer. Therefore, such models cannot be used for optimization of the structure and composition of ferroelectric polymer materials.

To address this limitation, we develop an analytical model describing the polymer phase transition from the crystalline to the condis phase, driven by the competition between van der Waals (vdW) inter-chain interactions and the conformational entropy contributions of gauche defects. All model parameters—including defect size and energy, lattice spacing, and interaction constants— are physically interpretable. To develop this approach, polyethylene was chosen as an initial modeling object as the simplest polymer having the condis phase. The external pressure for polyethylene is an analogue of the electric field for ferroelectric polymers, since the phase transition temperature depends on the pressure according to the Clapeyron-Clausius equation and as it was observed in MD simulations in current work. As will be shown below, the predictions of the proposed model are in perfect agreement with the results of MD simulations.

The structure of the paper is as follows. Section 2 contains the key observations from molecular dynamic modeling of the perfect polyethylene crystal. Here we used the same force-field, initial atomistic structure and simulation procedure as in [28]. The novel MD results are simulations at high external pressure (500 atm and 1000 atm). Additionally, the conformational entropy of the polymer chains in the polyethylene crystal was calculated as a function of temperature, using the equation derived in [29] for conformational entropy of polymer chain in vinylidene difluoride-based copolymers. In Section 3 we developed analytical model of polyethylene crystal, which describes the phase transition from monoclinic to condis phase. The model parameters possess distinct physical interpretations, including lattice parameter and additional energy of formation of gauche dihedral. This distinguishes presented model from phenomenological Landau-type models for VDF-based copolymers presented in [30] and [31]. Section 4 contains the model calibration and validation through MD simulations with different values of the pressure. Furthermore, Section 4 illustrates the sensitivity analysis of the model with respect to its key parameter—the average energy of gauche conformations. The simplified analytical model that captures the root cause of the first order of the phase transition to the condis phase is presented in Section 5. Discussion and conclusions are presented in Section 6.

The following two conventions are used throughout the paper. First, the term 'monomer' here means the minimal repeating unit of polyethylene, i.e. methylene group $CH_2$. All molar quantities here refer to mole of methylene groups. Second, we use the dimensionless molar entropy $S[\text{dimensionless}] = S[\text{J K}^{-1}\text{ mol}^{-1}]/R[\text{J K}^{-1}\text{ mol}^{-1}]$, measure molar energy in Kelvins $E[\text{K}] = E[\text{J mol}^{-1}]/R[\text{J K}^{-1}\text{ mol}^{-1}]$ and measure pressure in Kelvins per cubic meter $P[\text{K m}^{-3}] = P[\text{J m}^{-3}] \cdot N_A[\text{mol}^{-1}]/R[\text{J K}^{-1}\text{ mol}^{-1}]$, where $R$ is the molar gas constant and $N_A$ is Avogadro constant.

## 2. OBSERVATIONS FROM MOLECULAR DYNAMICS MODELING OF POLYETHYLENE CRYSTAL

As was noticed in the Introduction, polyethylene was chosen as an initial modeling object because it is the simplest polymer possessing the condis phase. The analytical model presented in this paper is based on the results of molecular dynamic (MD) simulation of the polyethylene crystal. Therefore, it is convenient to first present the major observations from MD modeling of polyethylene crystal. The results of MD simulations, that were used to calibrate/validate the analytical model, are presented in Section 3, where the analytical model is described.

The MD model of polymer crystal used in our simulations consists of 64 polymer chain in periodic boundary conditions. Each chain contains 100 methylene groups ($CH_2$). Initially the chains were packed in perfect monoclinic crystal. Further details of MD simulation can be found in Supplemental Material [32] (see also references [33–40] therein).

Formation of defects and transition into the condis phase were observed under the heating of this crystal. The visualization of crystal to condis crystal transition with rapid growth of the number of defects is presented in Figure 1. The defects were found short-living both in crystalline and in condis polyethylene. According to the histograms of gauche dihedral life duration (see details in Supplemental Material [32]), in crystalline phase at $T = 475$ K 99% of defects live less than 6.7 ps, in condis phase at $T = 485$ K 99% of defects live less than 16 ps. More than 1/3 of them live less than 20 fs in both phases. The average life durations are 57 and 482 fs respectively.

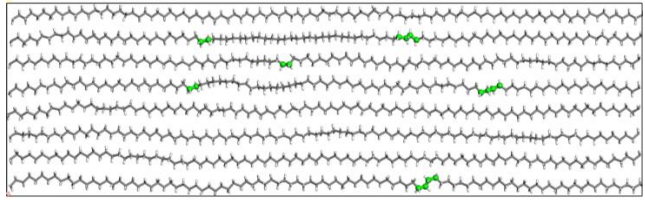

T = 477 K, $n_g$ = 1.3%

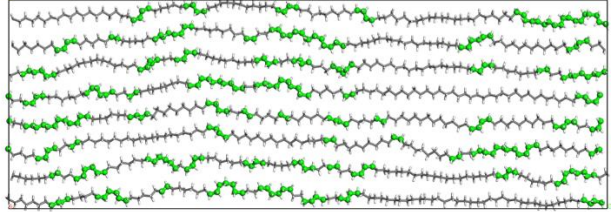

T = 484 K, $n_g$ = 20.4%

Figure 1. Visualization of one layer of polyethylene chains at temperatures slightly below (477K) and above (484K) the transition to the condis phase (MD snapshots). Carbon atoms are shown in grey, hydrogen atoms—in white. Central atoms (i.e. atoms $C^{(2)}$ and $C^{(3)}$ of $C^{(1)}$–$C^{(2)}$–$C^{(3)}$–$C^{(4)}$ bonded sequence) of *gauche* dihedrals are marked with green beads. A fraction of *gauche* dihedrals in the whole atomistic system, $n_g$, is specified for each snapshot.

As will be shown in the next Section, our analytical model is calibrated in such a way as to correctly describe the behavior of the polyethylene crystal in MD simulations at pressure 1 atm (which is unessentially different from zero for solid phase and can be set to zero in analytical calculations). For model validation, we compare the model predictions for the case of applied external pressures of 500 and 1000 atm with MD simulation data. Therefore, we performed MD simulations for three values of the external pressure (1, 500 and 1000 atm).

The temperature dependence of a polyethylene crystal density for three values of the external pressure (1, 500 and 1000 atm) are presented in Figure 2 (a). The temperature dependence of the crystal potential energy for the same three values of external pressure are presented in Figure 2 (b). It is seen from Figure 2 that both the density and the potential energy of the crystal undergo a discontinuous change at a certain temperature that depends on the value of the external pressure.

That indicates a first-order phase transition of the crystal at a certain temperature. This phase transition in the polyethylene crystal was previously studied by MD approach in [28,41] and it was identified as the phase transition from monoclinic to condis phase. The temperature of this phase transition depends on the value of the external pressure in accordance with the Clapeyron-Clausius equation:

$$\frac{dT_c}{dP} = \frac{T_c \Delta V}{\Delta H} \qquad (1)$$

Here $\Delta V$ and $\Delta H$ are the change of molar volume and enthalpy at the phase transition, respectively. Taking the values of $\Delta V \approx 2\text{Å}^3$ and $\Delta H \approx 300K$ per monomer from MD simulations for 1 atm pressure, the calculated changes in the phase transition temperature using Eq. (1) for 500 and 1000 atm are $12K$ and $24K$, respectively, which is in a perfect accord with the phase transition temperatures observed in the MD simulations.

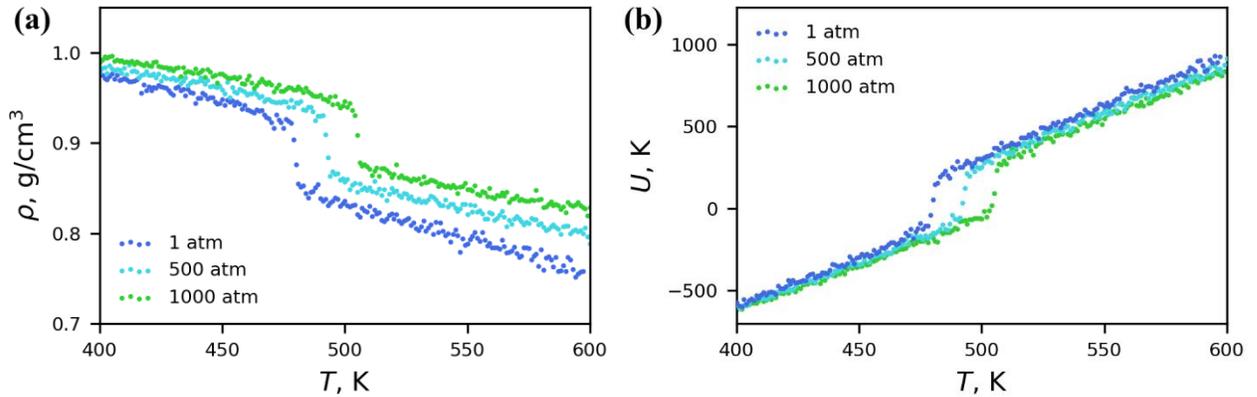

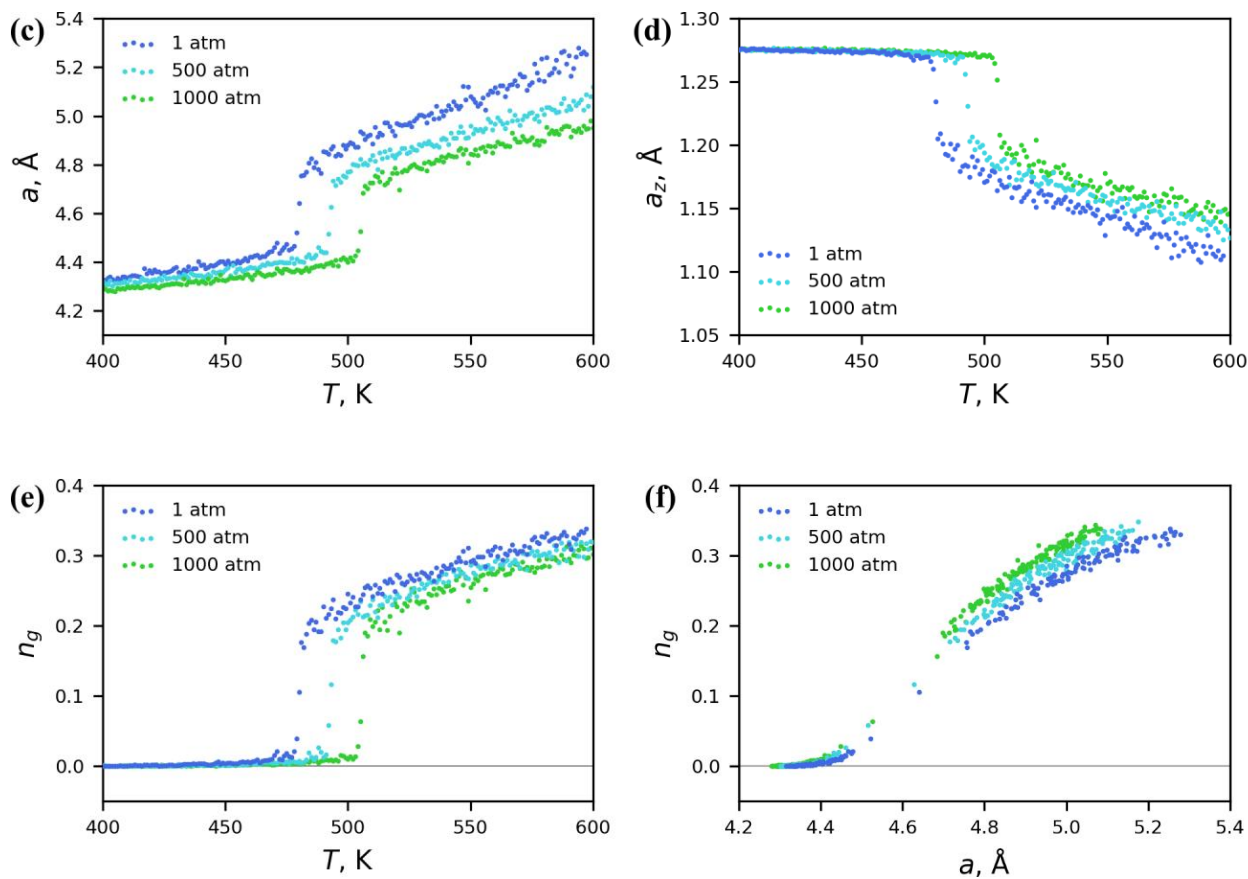

Figure 2. Results of MD simulation of polyethylene crystal heating at three values of external pressure (1, 500, and 1000 atm). (a) Density vs. temperature, (b) potential energy (per methylene group) vs. temperature, (c) transverse and (d) longitudinal lattice parameters vs. temperature, (e) fraction of *gauche* dihedrals vs. temperature and (f) vs. transverse lattice parameter.

The conformation of the polyethylene chain is determined by the sequence of its backbone C–C–C–C dihedrals. In saturated hydrocarbons (to which polyethylene belongs), there are three values of C–C–C–C dihedral achieving energetic minima and designated as *trans*, *gauche$^+$* and *gauche$^-$* (or $t$, $g^+$, and $g^-$). In the polyethylene condis phase, these minima are strongly dependent on the environment (i.e. configuration of neighboring monomers and neighboring chains), leading to the wide corresponding maxima on the dihedral population histogram. In addition, the temperature causes the fluctuations of dihedral angles near its equilibrium values. For these reasons, we treat these designations in a broad sense: as three intervals of a dihedral angle: *gauche*$^+$ (0°, 120°), *trans* (120°, 240°), and *gauche*$^-$ (240°, 360°).

The drastic increase of the number of *gauche* dihedrals in the transition from monoclinic to condis phase leads to the shrinking of the longitudinal *c* crystal dimension and to the growth of two transversal dimensions, ***a*** and ***b***. Let us introduce the "transverse" lattice parameter $a = \sqrt{ab}$ and "longitudinal" lattice parameter $a_z = c$, where ***a***, ***b*** and ***c*** (in a bold typeface) are lattice parameters of the crystal. Although there is no long-range order in the longitudinal direction above the phase transition temperature, by the longitudinal lattice parameter in condis phase we mean the average length of the chains per methylene group in the *c*-axis direction. The transverse and longitudinal lattice parameters as functions of temperature are presented in Figure 2 (c) and (d) for three values of external pressure.

The *gauche* dihedral fractions calculated from MD trajectories as functions of temperature are presented in Figure 2 (e) for three values of external pressure. The fraction of *gauche* dihedrals is close to zero in the monoclinic phase and undergoes a discontinuous change at the phase transition temperature. The fraction of *gauche* dihedrals in the condis phase monotonically increases with the increase in temperature. The *gauche* dihedrals in the monoclinic phase belong to rare defects, while there is a significant amount of the *gauche* dihedrals in the condis phase. The fractions of *gauche* dihedrals as functions of the transverse lattice parameter are presented in Figure 2 (f) for three values of the external pressure.

From comparison of Figures 2 (e) and (f) it is seen that, though the fraction of *gauche* dihedrals undergoes a discontinuous change as a function of the temperature, it is a smooth function of the transverse lattice parameter. That indicates that the average energy of formation of the *gauche* dihedral is a smooth function of the transverse lattice parameter. We utilized this observation in the analytical model of the polyethylene crystal.

The discontinuous change of the fraction of *gauche* dihedrals at the phase transition leads to discontinuous change the conformational entropy of polymer chain. The analytical equation for the dimensionless conformational entropy of the polymer chain was derived in [29]:

$$S_{con} = -\int P_2(\varphi_1,\varphi_2)\ln\left(\frac{P_2(\varphi_1,\varphi_2)}{\sqrt{P_1(\varphi_1)P_1(\varphi_2)}}\right)d\varphi_1 d\varphi_2 \qquad (2)$$

Here $P_1(\varphi)$ is a distribution function of dihedral angles and $P_2(\varphi_1,\varphi_2)$ is a pair distribution function of two consecutive dihedral angles along the polymer backbone. The pair distribution functions of the dihedral angles in the polyethylene crystal were calculated from the MD trajectories and analyzed in [28].

It was shown in [29] that the conformational entropy of the polymer chain, $S_{con}$, makes the major contribution to the entropy change under the phase transition of VDF-based copolymers to the condis crystal. The conformational entropy of the polyethylene crystal per one methylene group as a function of temperature was calculated by equation (2) for several values of temperature and atmospheric pressure. The calculated conformational entropy as a function of temperature is presented in Figure 3. It is seen that the conformational entropy undergoes a discontinuous change at the phase transition temperature (approximately 480 K). The increase in conformational entropy in the condis phase relative to the monoclinic phase is due to the chaotic formation of *gauche* dihedrals in the polyethylene backbone.

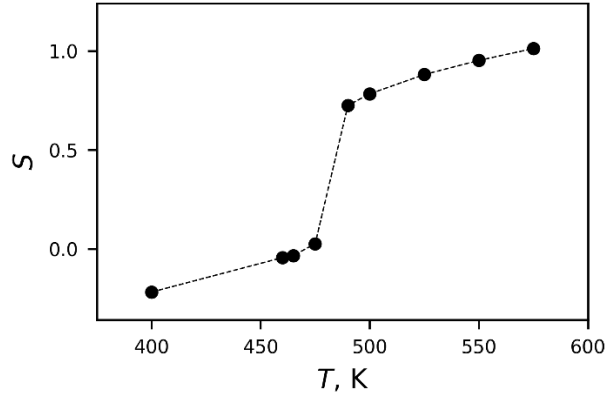

Figure 3. Dimensionless conformational entropy per dihedral angle as a function of temperature. (Dotted line segments are guide for the eyes).

The value of the calculated change of the conformational entropy at the phase transition is approximately 0.69. The total change of the entropy of the system at the phase transition in MD simulations, $\Delta S_{MD}$, is calculated from the change of the total energy at the phase transition, $\Delta U_{MD}$, as

$$\Delta S_{MD} = \frac{\Delta U_{MD}}{T_c} \quad (3)$$

The value of $\Delta U_{MD} = 344$ K and the phase transition temperature $T_c = 480$ K. Substituting these values into equation (3), we obtain $\Delta S_{MD} = 0.71$. That means that the major contribution to the entropy change at the phase transition of the polyethylene crystal to the condis phase also comes from the change in the conformational entropy as in the case of VDF-based copolymers. Moreover, the change of the conformational entropy is the driving force of the phase transition of polymer crystal to the condis phase and we believe that this is correct for all polymers possessing the condis phase.

## 3. ANALYTICAL MODEL

Typically, the crystalline phase of polymer has long-range order of setting angles along the chain, which disappears at the transition to the condis phase (the setting angle is the dihedral angle between the chain plane and the crystalline *ac* plane, assuming the chain is directed along the crystalline *c* axis). Therefore, the phase transition to the condis phase must be conventional second-order phase transition, accompanied by the vanishing of the long-range order and described by Landau theory of phase transitions [42]. However, in polyethylene and VDF-based copolymers, the first-order transition to the condis phase with discontinuous change in density and other physical quantities is experimentally observed.

To describe the first-order phase transition, in this section we present an analytical model of the phase transition of a polyethylene crystal from the monoclinic phase to the condis phase. Following Landau theory of phase transitions [42] the model is based on a construction of the free energy functional. However, the free energy of our model is written as a functional of the transverse lattice parameter of the crystal, rather than a long-range order parameter, as in Landau theory. The value of the transverse lattice parameter for given temperature and pressure is calculated from the minimization of the Gibbs free energy, $G(a, T, P)$. The analytical form of the Gibbs free energy captures the major physics that governs the formation of the condis phase, i.e. competition between conformational entropy of the chains and inter-chain vdW interaction that includes the additional energy of dihedrals (defects) formation. The model relates the fraction of *gauche* dihedrals in the crystal and conformational entropy with temperature and lattice parameter for given value of pressure. The model is calibrated against MD simulations of the polyethylene crystal at atmospheric pressure. Then the model allows to predict the temperature dependencies of the lattice parameter and the fraction of *gauche* dihedrals for arbitrary value of the pressure.

The total potential energy of the polymer chain in the crystal consists of two contributions: the internal energy of the chain and the energy of van der Waals (vdW) interactions with the surrounding chains. The total potential energy and vdW energy of the MD model of polyethylene crystal for three values of the pressure are plotted against the transverse lattice parameter, $a$, in Figure 4 (a) and (b), correspondingly. It should be noted that vdW energy in Figure 4 (b) consists of two contributions, i.e. inter-chain and intra-chain vdW interactions. It is seen from Figure 4 that variation of the total potential energy is approximately four times larger than variation of the vdW energy for the same interval of $a$. The sensitivity of the total potential energy to the pressure is also much higher than that of the vdW energy.

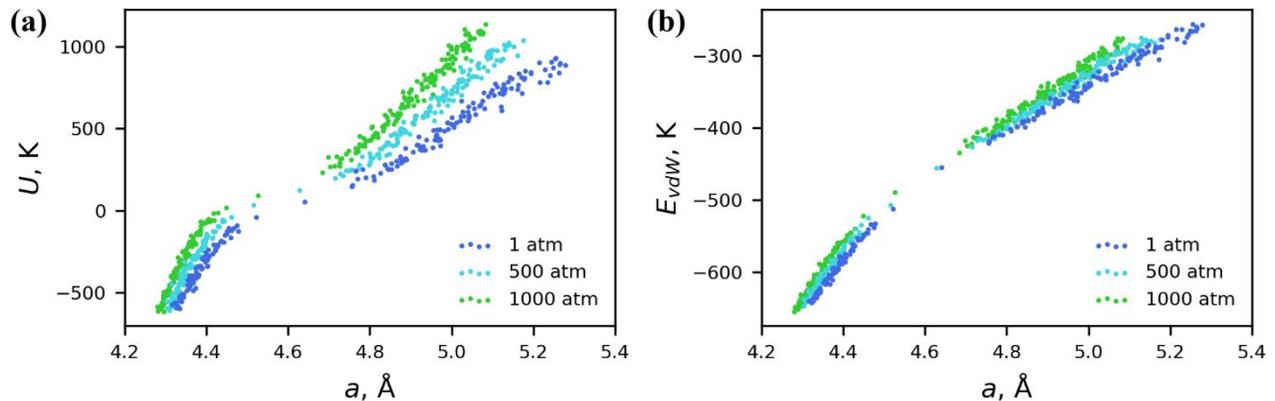

Figure 4. The total potential energy (a) and van der Waals energy (b) as functions of transverse lattice parameter for three values of the pressure from MD simulations.

The energy of vdW inter-chain interactions results in an effective tube-like potential. This tube-like potential prevents the given chain from moving far away from the axis of the "tube" [43]. The polymer chain has a large number of conformations in the tube that contributes into the conformational entropy of the chain. The value of the lattice parameter of the polymer crystal is determined by competition of the vdW inter-chain interaction and conformational entropy. The internal energy of the chain consists of the energy of the valence bonds, valence angles and dihedral angles which can be considered as internal degrees of freedom of the chain. It follows from Figure 4 that the internal energy of the chain is significantly greater than the vdW energy of inter-chain interaction. However, it is the inter-chain vdW interactions and the conformational entropy that determine the value of the lattice parameter, and they should be extracted from the total potential energy and total entropy in the model.

Based on the above, we write the Gibbs free energy of the polymer crystal per one methylene group as a sum of three contributions:

$$G(a,T,P) = F_{int}(T) + F_{el}(a,T) + PV(a,T) \qquad (4)$$

The first term, $F_{int}$, on the right hand side of this equation is the free energy associated with relative motion of the atoms in the chains, i.e. variation of the valence bonds, valence and dihedral angles. The second term, $F_{el}$, is an elastic free energy that is associated with chain motion in the tube. The third term is contribution of the pressure $P$, where $V$ is the volume per one methylene group.

Elastic free energy

The elastic free energy contains the elastic energy, $U_{el}$, of the vdW inter-chain interaction and conformational entropy, $S_{con}$, that is determined by the number of the chain conformations in the tube. Both elastic potential energy and conformational entropy depends on the temperature. The elastic energy and conformational entropy increase with temperature even when the volume of the crystal (transverse lattice parameter) is fixed. However, the explicit impact of temperature on the elastic energy and conformational entropy in polymer crystal is small and neglected in our approximation. We assume that the elastic energy and the conformational entropy are functions only of the transverse lattice parameter. So, the elastic free energy is written as follows

$$F_{el}(a,T) = U_{el}(a) - TS_{con}(a) \qquad (5)$$

The simplest structural element of polyethylene chain is methylene group (CH$_2$) that consists of three atoms. Therefore, each methylene group has 9 degrees of freedom. In the real polymer at temperature below 600 K the valence bonds and valence angles are partially frozen due to quantum effects. However, in the classical molecular dynamics all degrees of freedom are unfrozen and contribute into the total potential energy. Therefore, each internal degree of freedom can be approximately considered as a classical oscillator with parabolic potential for the treatment of MD results, though the actual inter-chain interactions (and the interactions in COMPASS force field) are more complicated. Thus, the internal free energy of the chain is approximated by the free energy of 9 independent oscillators

$$F_{int}(T) = \frac{n}{2}T - \frac{n}{2}T\ln\left(\frac{T}{T_0}\right) \tag{6}$$

Here $n=9$ is the number of oscillators, $T_0$ is some arbitrary temperature (the entropy in the classical thermodynamics is determined up to an arbitrary constant). The first and the second terms on the right hand side of equation (6) are the energy and the entropy of $n$ oscillators, respectively. The total potential energy and the total entropy of the crystal per one methylene group are calculated in our approximation as follows:

$$S(a,T) = S_{con}(a) + \frac{n}{2}\ln\left(\frac{T}{T_0}\right)$$
$$U(a,T) = U_{el}(a) + \frac{n}{2}T \tag{7}$$

It is seen from equation (7) that in our approximation both the energy and the entropy of the crystal depend separately on the lattice parameter (through inter-chain interaction) and on the temperature (through the internal energy of the chain, approximated by 9 oscillators).

The total potential energy of the crystal is known from quasi-equilibrium MD trajectory on gradual heating as a function of temperature, $U_{MD}(T)$. The total entropy of the crystal at MD trajectory can be calculated from well-known thermodynamic relation

$$\frac{\partial U}{\partial T} = T \frac{\partial S}{\partial T} \quad (8)$$

So, the total entropy at MD trajectory is calculated as

$$S_{MD}(T) = \int_{T_0}^{T} \frac{1}{T} \frac{\partial U_{MD}(T)}{\partial T} dT \quad (9)$$

Dependence of the lattice parameter on temperature is also the known function of temperature, $a_{MD}(T)$ (see Figure 2 (c)). The temperature can be considered as a function of the lattice parameter $T=T_{MD}(a)$ at the MD trajectory, where $T_{MD}(a)$ is inverse function of $a_{MD}(T)$. The total potential energy at MD trajectory, $U_{MD}(a)$, is also known from MD simulations (see Figure 4) and $S_{MD}(a)$ can be calculated form equation (9) after changing of the integration variable from $T$ to $a$. Substituting these functions into the left hand sides of equations (7) and substituting $T$ by $T_{MD}(a)$, we obtain the equations for elastic energy and conformational entropy:

$$U_{el}(a) = U_{MD}(a) - \frac{n}{2} T_{MD}(a)$$
$$S_{con}(a) = S_{MD}(a) - \frac{n}{2} \ln\left(\frac{T_{MD}(a)}{T_0}\right) \quad (10)$$

The polynomial splines of $U_{MD}(a)$ in monoclinic and condis phases for atmospheric pressure are presented in Figure 5 (a). The calculated $U_{el}(a)$ for monoclinic and condis phases are presented in the same figure. The total entropy, $S_{MD}(a)$, in monoclinic and condis phases are presented in Figure 5 (b). The conformational entropy, $S_{con}(a)$, calculated by equation (10) for monoclinic and condis phases are presented in the same figure. It is seen from Figure 5 that the variation of the total energy and total entropy is significantly greater than the variation of the elastic energy and the conformational entropy in the considered interval of *a*. That is due to the fact that the main contribution into the total energy and total entropy comes from valence bonds and valence and dihedral angles.

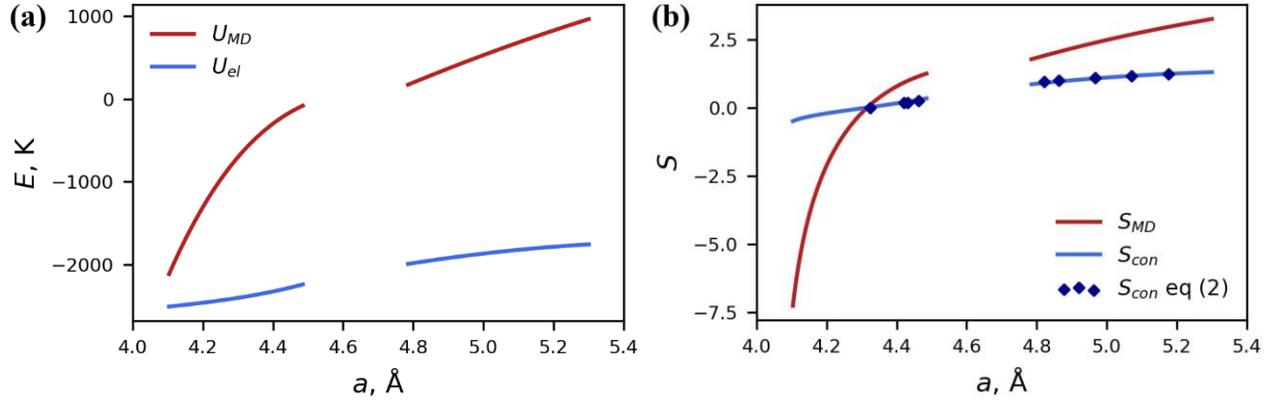

Figure 5. a) functions $U_{MD}(a)$ (red curves) at atmospheric pressure and $U_{el}(a)$ (blue curves); b) functions $S_{MD}(a)$ (red curves), $S_{con}(a)$ (blue curves) and conformational entropy calculated by equation (2) (symbols).

The conformational entropy, calculated by equation (2) for several values of temperature is shown in Figure 5 by symbols. It is seen that the conformational entropy, calculated directly from distribution of dihedrals, almost coincides with conformational entropy, calculated by the second equation (10), which was derived under assumption that the internal degrees of freedom of the chain can be approximated by classical oscillators. Therefore, subtracting the energy of the oscillators from the potential energy obtained in the MD simulation gives the elastic energy. As mentioned previously, the lattice parameter *a* at a specified temperature and pressure is determined through the minimization of the Gibbs free energy, given by Eq. (4), with respect to *a*. Since $F_{int}$ does not depend on *a* (see Eq. (6)), the lattice parameter $a(T)$ at zero pressure is calculated by minimizing the free energy functional (5), where elastic energy and conformational entropy are given by Eqs. (10). The Gibbs free energy for arbitrary pressure contains the volume per one monomer, that depends on the fraction of *gauche* dihedrals. Therefore, the calculation of *a(T)* for arbitrary pressure is required the knowledge of the fraction of gauche dihedrals, $n_g(a,T)$, that is calculated in the next Section.

### Contribution of *gauche* dihedrals

As it was mentioned earlier, the driving force of condis phase formation is increase of conformational entropy due to formation of *gauche* dihedrals. For calculation of the fraction of *gauche* dihedrals in the polymer chain we utilize the simple two-level model in which the *trans* and *gauche* dihedrals are distributed in the chain independently. This approximation is justified by a relatively small deviation of the probabilities of the pairs of the nearest dihedral angles such as *trans-trans, trans-gauche* and *gauche-gauche,* calculated from MD simulations from the corresponding product of probabilities of *trans* and *gauche* conformations. As it is seen in Figure 7 the accuracy of such approximation is not lower than 5%.

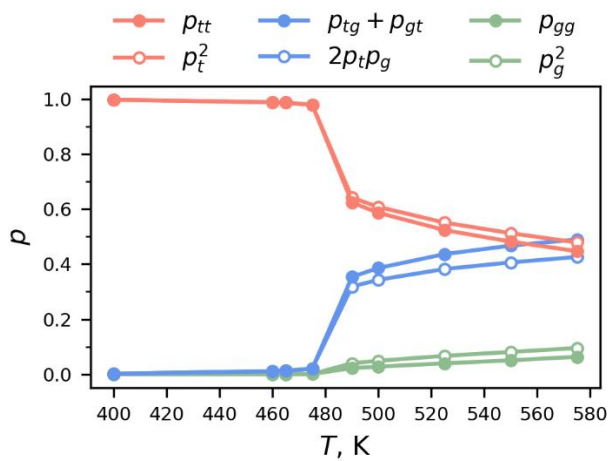

Figure 6. The probabilities of the pairs of the nearest dihedral angles *trans-trans,* $p_{tt}$, *trans-gauche,* $(p_{tg} + p_{gt})$, and *gauche-gauche,* $p_{gg}$, calculated from MD simulations and product of the corresponding probabilities of *trans* and *gauche* conformations, $p_t^2, 2p_t p_g, p_g^2$ as functions of temperature.

The partition function of such model per one monomer is

$$Z = \exp\left(-\frac{G_0(a,T)}{T}\right)\left(1+\exp\left(-\frac{\Delta G(a,T)}{T}\right)\right)$$
$$G_0(a,T) = U_0(a) - S_0(a)T + PV_0(a) \qquad (11)$$
$$\Delta G(a,T) = \Delta U(a) - T\Delta S(a) + P\Delta V(a)$$

Here, $U_0$, $S_0$ and $V_0$ are the energy, entropy and volume per one monomer of *trans* dihedral, respectively; $\Delta U$, $\Delta S$ and $\Delta V$ are the difference in the energy, entropy and volume between *gauche* and *trance* dihedrals, respectively. The entropy $S_0$ is associated with variation of the dihedral angles near their optimal values, i.e. the local conformational entropy. The Gibbs free energy per one monomer is calculated from thermodynamic relation $G = -T \ln Z$:

$$G(a,T) = G_0(a,T) + G_g(a,T)$$
$$G_g(a,T) = -T\ln\left(1+\exp\left(-\frac{\Delta G(a)}{T}\right)\right) \qquad (12)$$

Here $G_0$ is Gibbs free energy of *trans* dihedrals (ground state) and $G_g$ is additional contribution into Gibbs free energy from *gauche* dihedrals.

The local conformational entropies for *gauche* and *trans* dihedrals are estimated from histograms of dihedral angles obtained from MD trajectories. These estimates indicate that the difference between the local conformational entropy in *gauche* and *trans* conformation is small and $\Delta S$ can be taken equal to zero.

The fraction (concentration) of *gauche* dihedrals is calculated as follows:

$$n_g(a,T,P) = \frac{\partial G_g}{\partial \Delta U} = \frac{\exp\left(-\dfrac{\Delta U(a) - P\Delta V}{T}\right)}{1 + \exp\left(-\dfrac{\Delta U(a) - P\Delta V}{T}\right)} \tag{13}$$

Equation (13) contains unknown function $\Delta U(a)$ – the additional energy of *gauche* dihedrals relative to the energy of *trans* dihedrals. This function can be extracted from MD simulations if we know $n_g$ at MD trajectory. The energy of *gauche* dihedrals can be expressed through their concentration from equation (13) as follows:

$$\Delta U(a) = T \ln\left(\frac{1-n_g}{n_g}\right) + P\Delta V \tag{14}$$

This equation is used for calculation of $\Delta U(a)$ from MD trajectories. The difference between the volume of *gauche* and *trans* dihedrals is calculated for fixed value of $a$ and is equal to

$$\Delta V = a^2\left(a_{z,g} - a_{z,t}\right) = -a^2 \Delta a_z \tag{15}$$

Here $\Delta a_z = a_{z,t} - a_{z,g} = 0.41\,\text{Å}$, where $a_{z,t} = 1.277\,\text{Å}$ is a longitudinal size of monomer with all *trans* dihedrals and $a_{z,g} = 0.861\,\text{Å}$ is a longitudinal size of monomer with the *gauche* dihedral. The values of $a_{z,t}$ and $a_{z,g}$ follow from linear approximation of $a_z(n_g)$ obtained from MD simulations (see Figure 9 in the next Section).

The $n_g(T)$ and $a(T)$ are taken from MD trajectory for given value of pressure. Substituting these values of $n_g$ and $a$ into equation (14) we calculate the value of $\Delta U$ for this temperature and pressure. The value of $a$ for this temperature is plotted along the abscissa axis and calculated value of $\Delta U$ for the same temperature is plotted along the ordinate axis. The obtained dependencies $\Delta U(a)$ calculated for two values of the pressure (atmospheric and 1000 atm) are shown in Figure 7 by points. The plots in Figure 7 are based on both MD calculation and two-level analytical model. The analytical model is based on two assumptions: the *gauche* dihedrals are distributed along the chain independently and the additional energy of *gauche* dihedral, *ΔU*, is a function of lattice parameter *a* only. It is seen from Figure 7 that the points calculated for two values of the external pressure lie on the same universal curve which supports the model assumptions.

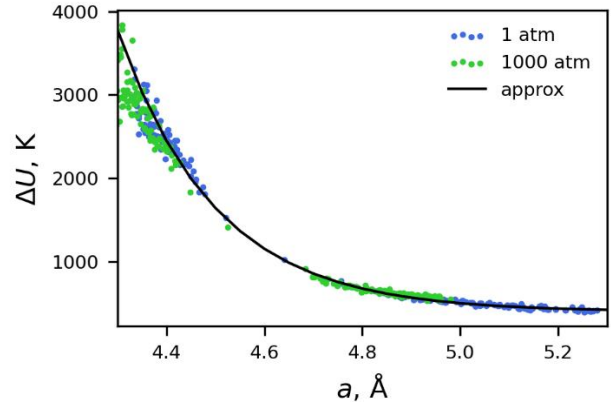

Figure 7. Dependence of the average energy of *gauche* dihedral on the lattice parameter. The points are calculated from MD simulations at the atmospheric pressure and at 1000 atm; solid line is an analytical approximation by equation (16).

This universal curve in Figure 7 can be approximated by the following analytical function:

$$\Delta U(a) = \Delta U_0 \exp\left(-\frac{a-a_0}{\Delta a_0}\right) + \Delta U_{00} \qquad (16)$$

Here $a_0 = 4.064$ Å is the value of the lattice parameter at zero temperature obtained from MD simulations. The values of other parameters in this equation are fitted to the points from Figure 7. The best set of parameter values are: $\Delta a_0 = 0.2$ Å, $\Delta U_0 = 11 \cdot 10^3$ K and $\Delta U_{00} = 400$ K. The function (16) with this set of parameter values is shown in Figure 7 by the solid line.

Now we can separate the contributions of the *gauche* dihedrals into elastic energy and conformational entropy:

$$U_{el}(a) = U_0(a) + U_g(a, T_{MD}(a))$$
$$S_{con}(a) = S_0(a) + S_g(a, T_{MD}(a))$$
(17)

Here $U_0$ and $S_0$ are elastic energy and the local conformational entropy of *trans* dihedrals, correspondingly; $U_g$ and $S_g$ are additional contributions into elastic energy and conformational entropy from *gauche* dihedrals, correspondingly. $U_g$ and $S_g$ for atmospheric pressure are calculated as follows (1 atm is unessentially different from zero for solid phase, so we take $P=0$ in the following equation):

$$S_g(a,T) = -\frac{\partial G_g}{\partial T} = \ln\left[1 + e^{-\frac{\Delta U(a)}{T}}\right] + \frac{\Delta U(a) n_g(a,T)}{T}$$
$$U_g(a,T) = G_g + TS_g = \Delta U(a) n_g(a,T)$$
(18)

Here $G_g$ is calculated by equation (12) and $n_g$ is calculated by equation (13). Both $S_g$ and $U_g$ vanish for $n_g=0$ and explicitly depend on $\Delta U(a)$. Therefore, the elastic energy and the local conformational entropy of *trans* dihedrals, $U_0(a)$ and $S_0(a)$, are calculated by the following equations:

$$U_0(a) = U_{MD}(a) - \frac{n}{2} T_{MD}(a) - U_g(a, T_{MD}(a))$$
$$S_0(a) = S_{MD}(a) - \frac{n}{2} \ln\left(\frac{T_{MD}(a)}{T_{in}}\right) - S_g(a, T_{MD}(a))$$
(19)

The functions $U_0(a)$ and $S_0(a)$, calculated from equation (19) are presented with a gap between monoclinic and condis phases in Figure 8 (a) and Figure 8(b), respectively. The total elastic energy and the total conformational entropy, $U_{el}(a)$ and $S_{con}(a)$, are presented in Figure 8 with a gap. As follows from Figure 8, $U_0$ coincides with $U_{el}$ and $S_0$ coincides with $S_{con}$ in the interval of monoclinic phase ($a < 4.4$Å), where there is a negligible fraction of *gauche* conformations. It is also seen in Figure 8 that $U_0(a)$ and $S_0(a)$ can be approximated by one smooth curve in both monoclinic and condis phases.

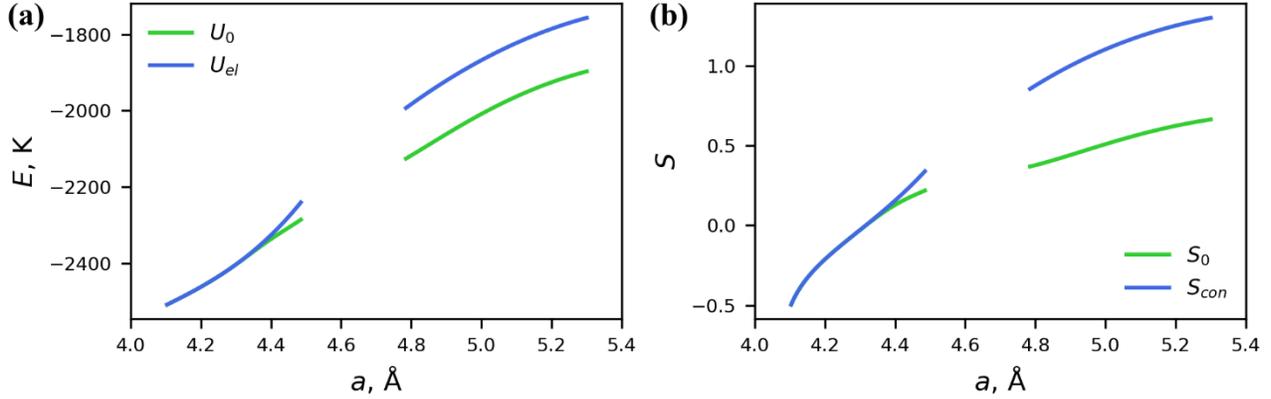

Figure 8. a) functions $U_{el}(a)$ for atmospheric pressure and $U_0(a)$ for MD model of polyethylene crystal; b) functions $S_{con}(a)$ and $S_0(a)$ for MD model of polyethylene crystal.

Impact of pressure

The volume of the system per one monomer, $V$, depends on the value of the lattice parameter in transverse and longitudinal directions: $V=a^2 a_z$. The formation of *gauche* dihedrals causes increase of $a$ and decrease of $a_z$ (see Figure 2 (d)). Therefore, the lattice parameter in the longitudinal direction should be dependent on the fraction of *gauche* dihedrals, $n_g$. The dependence of $a_z$ on $n_g$ obtained from MD simulations is presented in Figure 9 for two values of external pressure. It is seen from Figure 9 that dependences $a_z(n_g)$ for two values of pressure are practically identical. Therefore, we assume that $a_z(n_g)$ is a universal function of $n_g$, which can be approximated by the linear expression:

$$a_z(n_g) = a_{z,t} \cdot (1-n_g) + a_{z,g} \cdot n_g \qquad (20)$$

The fraction of gauche dihedrals is function of $a$ and $T$ and is calculated by equation (13).

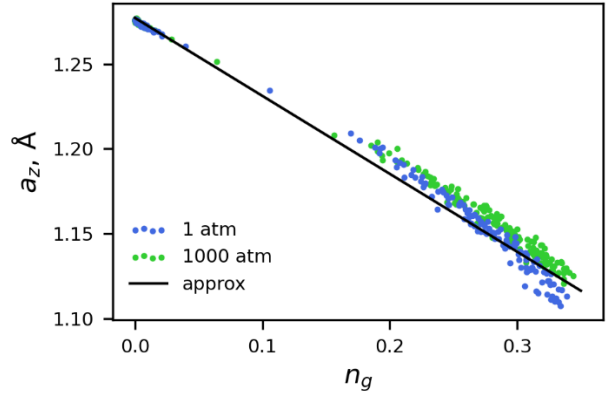

Figure 9. Dependence of longitudinal lattice parameter on the fraction of *gauche* dihedrals for two values of external pressure.

It should be noted that ignoring the dependence of $a_z$ on $n_g$ leads to a significant error in the phase transition temperature at elevated pressure. In this case the model would overestimate the shift of $T_c$ due to the pressure by a factor of about two.

Model overview

Finally, we obtained the closed system of equations for Gibbs free energy of the polymer crystal. The Gibbs free energy of the crystal is given by the following equation:

$$G(a,T) = U_{MD}(a) + \frac{n}{2}(T - T_{MD}(a)) - T\left(S_{MD}(a) + \frac{n}{2}\ln\left(\frac{T}{T_{MD}(a)}\right)\right) + Pa^2 a_z(n_g(a,T)) \quad (21)$$

Since the lattice parameter $a_{MD}(T)$ undergoes discontinuous change at the phase transition temperature, $T_c$, the function $T_{MD}(a)$ is a continuous function at two intervals, in monoclinic and condis phases. The functions $T_{MD,m}(a)$ and $T_{MD,c}(a)$, $U_{MD,m}(a)$ and $U_{MD,c}(a)$ are taken as an approximation of MD data for transverse lattice parameter and potential energy with polynomials in monoclinic and condis phase respectively at atmospheric pressure in order to eliminate a noise, unavoidable in MD (see details in Supplemental Material [32]).

The total entropy is calculated by equation (9) with replacement of the integration variable

$$S_{MD,m}(a) = \int_{a_{in}}^{a} \frac{1}{T_{MD,m}(a)} \frac{dU_{MD,m}}{da} da$$
$$S_{MD,c}(a) = S_{MD,m}(a_m) + \frac{\Delta U_{MD}}{T_c} + \int_{a_c}^{a} \frac{1}{T_{MD,c}(a)} \frac{dU_{MD,c}}{da} da \quad (22)$$

Here $a_{in} = a_{MD,m}(T_{in})$ is some initial value of lattice parameter; $a_m$ and $a_c$ are the values of the lattice parameter at $T_c$ in monoclinic and condis phase, correspondingly; $\Delta U_{MD}$ is the energy change at the phase transition ($\Delta U_{MD} = \Delta U_{MD,c}(a_c) - \Delta U_{MD,m}(a_m)$). Substituting $U_{MD,m}(a)$, $S_{MD,m}(a)$ and $T_{MD,m}(a)$ into equation (21) we obtain the analytical function for the Gibbs free energy for MD model of polyethylene crystal in the monoclinic phase. Substituting $U_{MD,c}(a)$, $S_{MD,c}(a)$ and $T_{MD,c}(a)$ into equation (21) we obtain the analytical function for the Gibbs free energy for MD model of polyethylene crystal in the condis phase. The dependence of longitudinal lattice parameter, $a_z$, on the fraction of gauche dihedrals, $n_g$, is given by equation (20), where $n_g$ is calculated by equation (13) with $\Delta U(a)$ given by equation (16).

Minimization of the Gibbs free energy (21) for monoclinic and condis phases for given values of $T$ and $P$ results in the optimal values of $a$ in monoclinic and condis phases. The phase that provides the global minimum is the equilibrium phase for these values of $T$ and $P$. Varying the temperature at fixed $P$, we calculate the lattice parameter as a function of $T$ for a given value of $P$, $a(T, P)$. Substituting this $a(T, P)$ into equation (13) we obtain the fraction of *gauche* dihedrals as a function of temperature for given values of $P$, $n_g(T, P)$.

In the proposed approach, the calibration of the analytical model has been performed using molecular dynamics (MD) simulations of idealized polyethylene crystals. Although the MD approach provides a reasonably accurate representation of real systems, it must be acknowledged that the employed force fields inherently approximate complex interatomic interactions within real materials. While MD simulations offer valuable insights into atomic-scale phenomena, their limitations prevent precise predictions of macroscopic physical characteristics, particularly critical parameters like phase transition temperature. Consequently, direct comparisons between MD-simulated predictions and empirical observations frequently reveal substantial deviations.

Nevertheless, if our analytical model were calibrated directly against experimental data, its predictive capability would align more closely with observed measurements. A key challenge arises when attempting to compute elastic energies using experimental dependencies of system energy versus temperature combined with Equation (10); this is impractical since the internal degrees of freedom associated with polymer chains deviate significantly from classical harmonic behavior due to underlying quantum effects.

Fortunately, there is an alternate method. By experimentally determining both the lattice constant $a_{exp}(T)$ and the fraction of gauche dihedral angles $n_g(T)$ at specific temperatures $T_{exp}$, for single-crystal samples, one may employ Equation (14) under zero pressure conditions to estimate the energy contribution from gauche configurations relative to changes in the lattice parameter. Additionally, the entropic contributions arising from gauche formations can be quantified at each experimental point via Equation (23):

$$S_g(T_{exp}) = (n_g(T_{exp}) - 1)\ln(1 - n_g(T_{exp})) - \ln(n_g(T_{exp}))n_g(T_{exp}) \qquad (23)$$

This equation follows from equations (14) and (18). The local conformational entropy, $S_0$, (caused by thermal fluctuations in the local minima) can be approximated by the entropy of classical oscillator, $S_0(T) = \ln(\sqrt{T/T_0})$, where $T_0$ is arbitrary constant. The elastic energy can be calculated using thermodynamic relation (8) $\quad U_{el}(a) = \frac{1}{2}(T_{exp}(a) - T_0) + \int_{T_0}^{T_{exp}(a)} T dS_g(T) \qquad (24)$

Where $T_{exp}(a)$ is the inverse function for experimentally measured function $a_{exp}(T)$.

Finally, the total Gibbs free energy functional takes the following form

$$G_{\exp}(a,T,P) = U_{el}(a) - T\left(S_g\left(T_{\exp}(a)\right) + S_0\left(T_{\exp}(a)\right)\right) + Pa^2 a_z \qquad (25)$$

Here $U_{el}(a)$ is evaluated according to equation (24). The thermal expansion coefficient or elastic modulus can subsequently be extracted through minimization of Gibbs free energy (25). The model is validated by comparing the predicted crystal properties with the corresponding experimental measurements.

## 4. ANALYTICAL MODEL PREDICTIONS

In this section, the analytical model predictions for polyethylene crystal are presented and compared with the results of MD simulations. The analytical model described in the previous section was calibrated against MD data for atmospheric pressure. The calibrated model allows to predict the transversal lattice parameter, $a$, and the fraction of *gauche* dihedrals, $n_g$, as functions of temperature for an arbitrary value of the pressure. Also, we present in this Section the analysis of the model sensitivity to the key model parameters, $\Delta a$ and $\Delta U_0$, that govern the formation of condis phase.

The Gibbs free energies of monoclinic and condis phases as functions of lattice parameter at atmospheric pressure are shown in Figure 10 (a) at phase transition temperature $T_c=480$ C. At $T_c$ the Gibbs free energy of the monoclinic phase in the minimum ($a=a_m$) is equal to the Gibbs free energy of the condis phase in the minimum ($a = a_c$). At $T<T_c$ the minimum of the Gibbs free energy in the monoclinic phase is lower than the minimum of the Gibbs free energy in the condis phase, so that the monoclinic phase is an equilibrium state at $T<T_c$. At $T>T_c$ the minimum of the Gibbs free energy in the monoclinic phase is higher than the minimum of the Gibbs free energy in the condis phase, so that the condis phase is an equilibrium state at $T>T_c$. At $T=T_c$ the crystal transfers from the monoclinic to condis phase under the heating. That results in discontinuous increase of the lattice parameter from $a_m$ to $a_c$. The fraction of *gauche* dihedrals as a function of the lattice parameter at $T=T_c$ and atmospheric pressure is shown in Figure 10 (b). It is seen from Figure 10 (b) that the fraction of *gauche* dihedrals discontinuously increases from 0.03 to 0.19 at the phase transition, when the lattice parameter increases from $a_m$ to $a_c$.

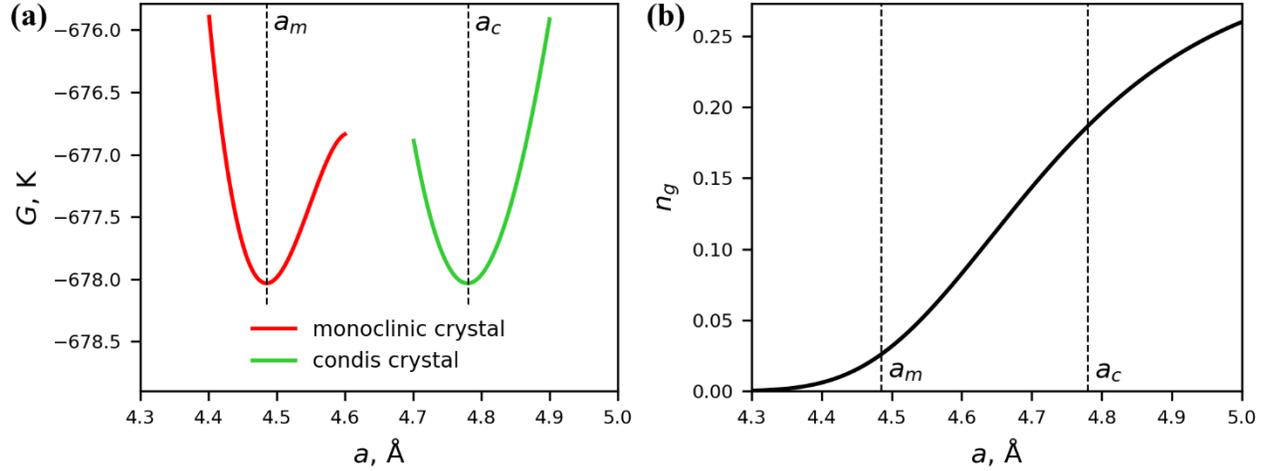

Figure 10 a) Gibbs free energies in monoclinic (red curve) and condis (green curve) phases as functions of lattice parameter at phase transition temperature $T_c$=480 C and atmospheric pressure, calculated by eq. (21); b) fraction of *gauche* dihedrals as a function of lattice parameter under the same conditions, calculated by eq. (13).

Varying the temperature and finding the global minimum of the Gibbs free energy, we obtain the lattice parameter as a function of $T$. Substituting this function $a(T)$ into equation (13), we calculate the fraction of *gauche* dihedrals as a function of temperature, $n_g(T)$. The calculated $a(T)$ and $n_g(T)$ for three values of external pressure (atmospheric pressure, 500 atm and 1000 atm) are presented in Figure 11 (a). The data from MD simulations are shown by the dots in the same figure. The model prediction for atmospheric pressure excellently agrees with MD data because the model was calibrated against these data. The model predictions for pressure of 500 atm and 1000 atm are also in a perfect agreement with MD data. That indicates that our model captures the key physics that governs formation of the condis phase.

However, the model agreement with MD data decreases with increase of the pressure. That is due to the fact that our approximation is an expansion of the Gibbs free energy near the MD trajectory at the atmospheric pressure, i.e. it is correct only in some vicinity of MD trajectory at the atmospheric pressure. At the atmospheric pressure, minimization of Gibbs free energy (21) results in $a(T)$, which coincides with MD trajectory $a_{MD}(T)$. However, when the pressure increases, the trajectory of the system moves away from the trajectory at atmospheric pressure, which results in decrease of the model accuracy.

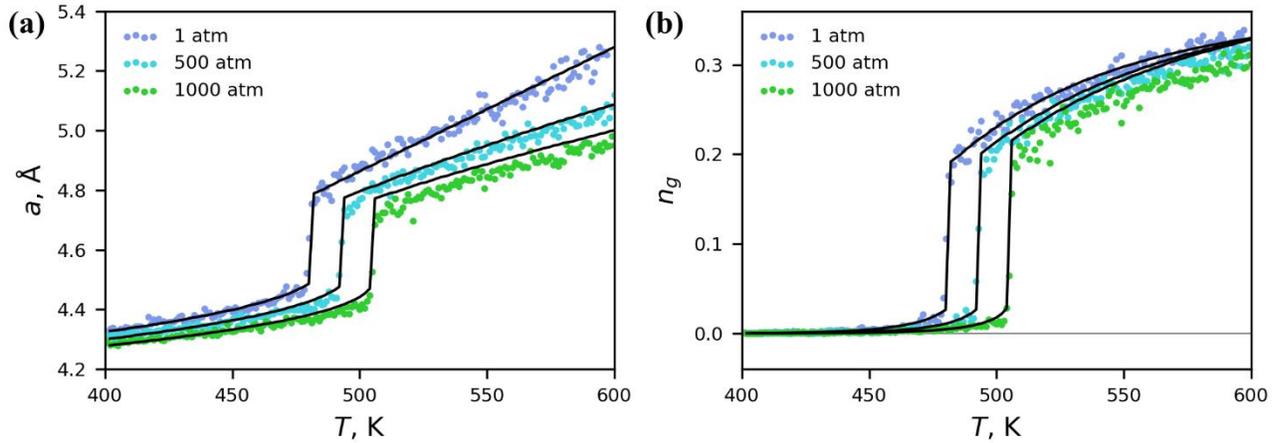

Figure 11. (a) Lattice parameter, calculated from minimization of eq. (21) with respect to $a$, and (b) fraction of gauche dihedrals, calculated by eq. (13), as functions of temperature for three values of the pressure. Solid lines – calculated by analytical model, points – MD simulation.

The additional energy of the crystal caused by the formation of *gauche* dihedrals is approximated in our model by equation (16) that contains two key parameters: $\Delta a$ – the typical size of "defect" with *gauche* dihedral, and $\Delta U_0$ – the typical additional energy of "defect" with *gauche* dihedral.

We believe that the functional form of *ΔU(a)* for all polymers with flexible chains and without side groups are close to that given by equation (16), but the values of *Δa* and *ΔU₀* vary for different polymers. Therefore, the details of the phase transition to condis phase for different polymers can be studied by variation of *Δa* and *ΔU₀* in our model. Here for simplicity we also assume that the additional energy of *gauche* dihedral is proportional to the typical size of "defect" with *gauche* dihedral:

$$\Delta U_0 = \frac{\Delta U_{0,ref}}{\Delta a_{ref}} \Delta a \tag{26}$$

Here $\Delta U_{0,ref} = 11000$ K and $\Delta a_{ref} = 0.2$ Å are the values of these parameters derived from MD model of polyethylene.

The calculated dependencies of the lattice parameter and the fraction of *gauche* dihedrals on the temperature for three values of *Δa* (0.2, 0.15 and 0.1 Å) are presented in Figures 12 (a) and (b), correspondingly. The phase transition temperature, $T_c$, drastically decreases with decrease of *Δa* as seen from Figure 12. The decrease of $T_c$ is accompanied by the decrease of discontinuous change of *a(T)* and $n_g(T)$ at $T = T_c$. Moreover, the discontinuous changes of *a(T)* and $n_g(T)$ at $T_c$ are not observed for *Δa*=0.1 Å; the change of *a(T)* in vicinity of $T_c$ remains fast but continuous. The $n_g(T)$ tends to zero while the derivative of $n_g(T)$ tends to infinity at $T = T_c$ for *Δa* = 0.1 Å. The small fraction of $n_g$ just below $T_c$ is comprised of the rare defects with *gauche* dihedrals in the monoclinic phase. Such behavior of *a(T)* and $n_g(T)$ indicates that there is no first-order phase transition to the condis phase for *Δa* = 0.1 Å. Therefore, the analytical model predicts that the phase transition of polymer crystal from 3D crystalline phase to the condis phase can be of the first-order for large value of *Δa* (and *ΔU₀*) and of the second or higher order for the small value of *Δa* (and *ΔU₀*). The physical mechanism underlying such dependence of the system's thermodynamic properties on defect size will be analyzed using a simplified model in the following section.

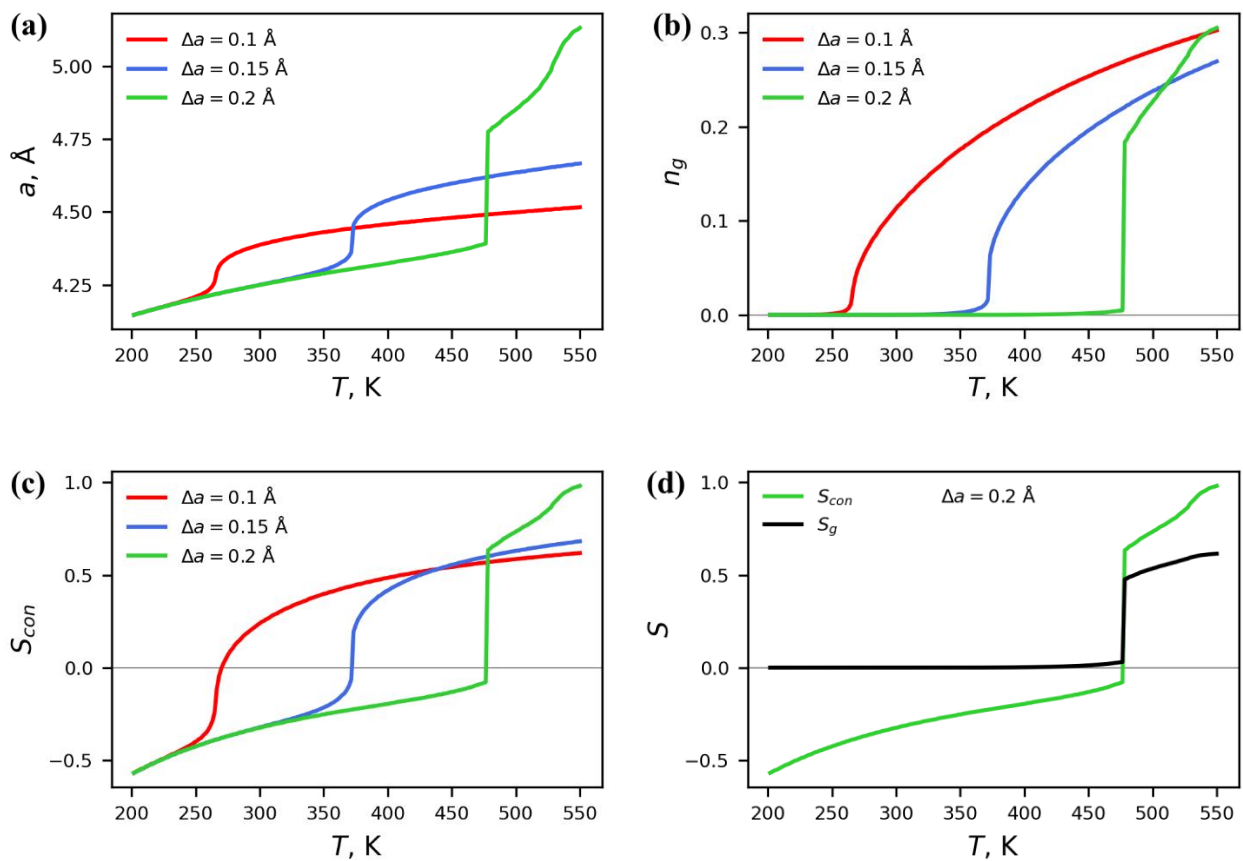

Figure 12. For three values of *Δa* (0.2, 0.15 and 0.1 Å): (a) the lattice parameter as a function of temperature, calculated from minimization of eq. (21) with respect to *a* with modified *ΔU(a)*, (b) the fraction of *gauche* conformations as a function of temperature, calculated by eq. (13) with modified *ΔU(a)*, (c) total conformational entropy as a function of temperature, calculated by eq. (17) and eq. (18) with modified *ΔU(a)*. For *Δa* = 0.2 Å: (d) total conformational entropy, calculated by eq. (17) and conformational entropy caused by formation of *gauche* dihedrals, $S_g$, calculated by eq. (18), as functions of temperature.

The calculated dependencies of the conformational entropy on the temperature for three values of $\Delta a$ (0.2, 0.15 and 0.1 Å) are presented in Figure 12 (c). The discontinuous change of the conformational entropy at $T = T_c$ is observed for $\Delta a$ = 0.2 Å and $\Delta a$ = 0.15 Å. That indicates the 1$^{st}$ order phase transition for these values of $\Delta a$. The entropy changes continuously in the vicinity of $T_c$ for $\Delta a$ = 0.1 Å, which indicates that the phase transition for this value of $\Delta a$ is of the 2$^{nd}$ order or higher.

The conformational entropy consists of two contributions: the local conformational entropy, $S_0$, caused by variation of the *trans* and *gauche* dihedral near their average values and the entropy caused by formation of *gauche* dihedrals, $S_g$. The potential energy of dihedrals has three local minima: one lowest minimum corresponds to *trans* dihedral (trans-minimum) and two energetically equivalent minima correspond to two *gauche* dihedrals (gauche-minimum). The energy difference between trans-minimum and gauche-minimum in the crystal depends on the lattice parameter and is equal to $\Delta U(a)$. In the monoclinic phase far from phase transition the majority of the dihedrals is located in trans-minima and conformational entropy is caused by thermal induced deviations of the dihedrals around the trans-minimum. The entropy increases with temperature due to increase of the deviations amplitude. The further increase of the temperature causes formation of significant fraction of *gauche* dihedrals. If all *trans* and *gauche* dihedrals would be located in the local minima, the conformational entropy would have calculated exactly by eq. (18). However, the thermal fluctuations cause deviation of the dihedrals from the local minima both for *trans* and *gauche* dihedrals. That results in additional conformational entropy, $S_0$. The total conformational entropy $S_{con}$ is equal to the sum of $S_0$ and $S_g$. The total conformational entropy, $S_{con}$, and the conformational entropy caused by formation of *gauche* dihedrals, $S_g$, as functions of temperature are presented in Figure 12 (d) for $\Delta a = 0.2$ Å (MD model of polyethylene crystal).

The discontinuous change of $S_{con}$ at $T_c$ is equal to 0.71, while the discontinuous change of $S_g$ at $T_c$ is equal to 0.45. Therefore, the contribution of the formation of *gauche* conformations into the entropy change at $T_c$ is approximately equals to 63% and other 37% comes from increase of the local conformational entropy (increase of the width of distribution of dihedral angles in *trans* and *gauche* local minima). The increase of the local conformational entropy at $T_c$ is caused by the increase of the lattice parameter that facilitates the variation of dihedral angles near equilibrium value. Therefore, we can conclude that the local conformational entropy and the entropy caused by formation of *gauche* dihedrals give the comparable contribution into the entropy change under the phase transition to condis phase.

## 5. SIMPLIFIED ANALYTICAL MODEL

To identify the root cause of the first order of the phase transition to the condis phase, in this section we present the simplest model, which includes only some general properties of polymer chains that play a key role in the phase transition. That is, the simplified model does not pretend to describe accurately any particular polymer, like the analytical model developed in the previous section. It is a minimal model that will help us understand the physics of the phase transition to the condis phase in the simplest way.

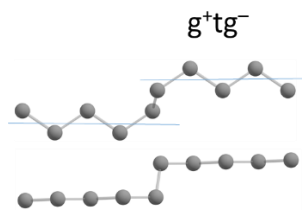

Figure 13. Two projections of the most common defect in polymer chains of the type $g^+tg^-$ observed in MD simulations.

Similar to the above analytical model, the simplified model describes one polymer chain in an imaginary potential tube formed by vdW interactions with neighboring chains [43]. A typical configuration of the polymer chain in a crystallite is as follows. The major part of the dihedral angles of the backbone is *trans* dihedrals, so the chain axis coincides with the tube axis, and the monomers are located most probably close to the center of the tube. Relatively rarely and randomly, defects containing one or a few *gauche* dihedrals are formed in the chain. The most frequently observed in MD simulations defect consists in the following sequence of dihedral angles along the chain: …tg$^+$tg$^-$t… or …tg$^-$tg$^+$t…. The form of the chain near such defect is shown in Figure 13. As it is seen in Figure 13, sequence tg$^+$tg$^-$t does not change the direction of the chain axis, but only shifts the axis by a certain distance, of the order of the length of the C-C bond. When the chain with defect tg$^+$tg$^-$t is in the potential tube, only a few monomers in the vicinity of the defect are significantly displaced from the center of the tube as shown in Figure 14 (a).

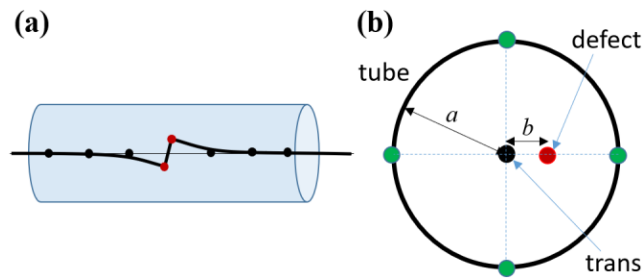

Figure 14. (a) Scheme of the defect of polymer chain in the potential tube formed by the neighboring chains. (b) Schematic diagram of the mutual arrangement of *regular* (black circle) and *defect* (red circle) monomers and neighboring chains (green circles) in the simplified model.

Such a defect is like a knot on polymer chain (or a local thickening of a chain), and does not cause neighboring chains to deform strongly. Therefore, the potential energy of such a defect is not too high and, as a result, it is much more common than defects of other types.

On the basis of the above picture, and for the sake of simplicity, we assume that the thermal fluctuations are relatively small, so that there are just two possible states that each monomer can occupy. The first 'ground' state is located in the center of the tube (black circles in Figures 14 (a) and (b)) it is occupied by 'regular' monomers, which do not belong to defects. The second, 'defect,' state is shifted from the center of the tube by a fixed distance $b$, which we will call the defect size (red circles in Figures 14 (a) and (b)) and which is directly related to the parameter $\Delta a$ in Eq. (26). It is assumed that the *defect* monomers are randomly distributed along the chain.

For simplicity, we also assume that the potential in the tube is formed by 4 neighboring chains (green circles in Figure 14 (b)) and the displacement of defect monomer is directed towards one of them. According to this assumption, the potential energy of the regular monomer depends on the tube radius (the lattice parameter) $a$ as:

$$U_0(a) = 4U_{LJ}(a) \tag{27}$$

Here $U_{LJ}(a)$ is a sum of vdW interactions of the regular monomer with all the monomers of the neighboring chain. For definiteness, we assume that this vdW interaction is described by the Lennard-Jones 9-6 potential (following COMPASS force field used for MD in this paper)

$$U_{LJ}(x) = \varepsilon \left[ 2\left(\frac{a_0}{x}\right)^9 - 3\left(\frac{a_0}{x}\right)^6 \right] \tag{28}$$

Here $a_0$ is the location of the minimum of $U_{LJ}(x)$, and $\varepsilon = -U_{LJ}(a_0)$ is the depth of the potential well. Therefore, $a_0$ sets the spatial scale and $\varepsilon$ sets the energy scale.

According to Figure 14 (b), the potential energy of the defect monomer with four neighboring chains is

$$U_d(a,b) = 2U_{LJ}\left(\sqrt{a^2+b^2}\right) + U_{LJ}(a-b) + U_{LJ}(a+b) \tag{29}$$

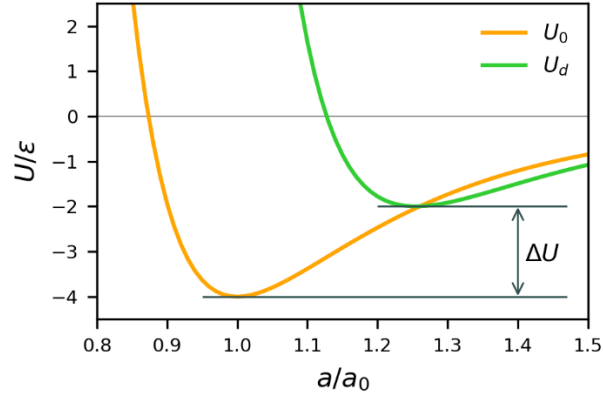

Figure 15. Dependence of potential energy for regular ($U_0$) and defect ($U_d$) monomers on the lattice parameter $a$ for defect size $b = 0.3a_0$.

In Figure 15 the potential energies for regular ($U_0$) and defect ($U_d$) monomers are shown as functions of the lattice parameter $a$ for fixed defect size $b = 0.3a_0$. As shown in Figure 15, the minimum of the potential energy for the regular monomer is achieved at $a_{min,0} = a_0$, and the potential energy smoothly increases for $a > a_0$. The potential energy of the defect monomer has a minimum at $a_{min,d} \approx a_0 + b$, where the value of the energy is higher than the energy of the regular monomer by the value $\Delta U$, which has the physical meaning of the defect energy given by Eq. (26) in the analytical model. The shift of $a_{min,d}$ to a higher value of the tube size $a$ is caused by the hard-core repulsion of the defect monomer from the nearest neighboring chain (the right green circle in Figure 14 (b)).

The partition function of a chain containing $N$ independent monomers is $Z^N$, where $Z$ is a partition function per monomer, which consists of regular and defect contributions:

$$Z(a,b,T) = \exp\left(-\frac{U_0(a)}{T}\right) + \exp\left(-\frac{U_d(a,b)}{T}\right) \qquad (30)$$

In the last equation, we neglected thermal fluctuations and assumed that the regular and defect monomers are located exactly in their optimal positions, as shown in Figure 14 (b). (We have verified that taking thermal fluctuations into account does not lead to a qualitative change in the results of the simplified model.)

The free energy of the chain is given by

$$F(a,b,T) = -T \ln Z \qquad (31)$$

The lattice parameter $a$ is a model parameter, which is determined by the minimum free energy condition. As an example, the dependence of the free energy on the lattice parameter $a$ for the defect size $b = 0.3a_0$ and three different temperatures is shown in Figure 16 (a).

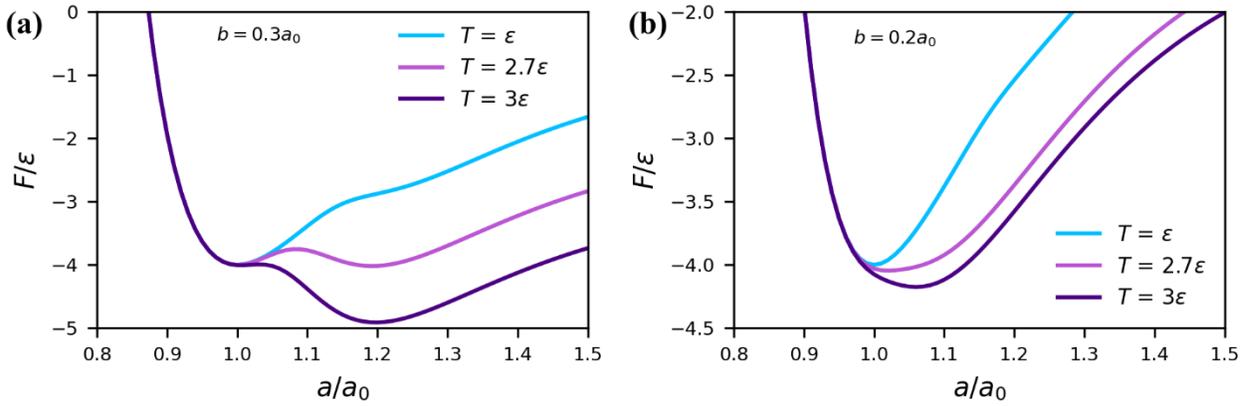

Figure 16. Dependencies of the free energy on the lattice parameter $a$ for defect sizes (a) $b = 0.3a_0$ and (b) $b = 0.2a_0$ and temperatures $T = \varepsilon$, $T = 2.7\varepsilon$, and $T = 3\varepsilon$.

As follows from Figure 16 (a), for the defect size $b = 0.3a_0$ the simplified model predicts two minima of the free energy and the first-order phase transition at $T \approx 2.7\varepsilon$. This phase transition temperature can easily be estimated as follows. The first minimum of free energy is located very close to $a \approx a_0$, where the partition function (30) is

$$Z_1 \approx \exp\left(-\frac{4\varepsilon}{T}\right) \qquad (32)$$

Here we neglect the contribution from defect monomers, $Z_d$, because as it follows from Figure 15, the energy of defect monomers, $U_d$ is very high at $a \approx a_0$ and the second term on the right hand side of equation (30) is exponentially small.

The second minimum of free energy in Figure 16 (a) occurs at $a \approx 1.2a_0$, where the potential energy of the regular monomers is close to that of the defect monomers and is equal to $U_0(1.2a_0) \approx U_d(1.2a_0) \approx -4\varepsilon + \Delta U$. Therefore, the regular and defect monomers give approximately equal contribution to the partition function (30) near the $a = 1.2a_0$, which is

$$Z_2 \approx 2\exp\left(-\frac{4\varepsilon + \Delta U}{T}\right) \qquad (33)$$

The phase transition occurs at temperature, when the partition functions $Z_1$ and $Z_2$ become equal, $Z_1 = Z_2$. This equation gives the estimate for the phase transition temperature as

$$T_C = \frac{\Delta U}{\ln 2} \qquad (34)$$

For the case under consideration, $b = 0.3a_0$, the value $\Delta U \approx 2\varepsilon$ (see Figure 15) and the estimate $T_C \approx 2.8\varepsilon$ agrees well with the exact value of $T_C = 2.7\varepsilon$ presented in Figure 16 (a).

The dependence of the free energy on the lattice parameter $a$ becomes qualitatively different for the smaller defect size $b = 0.2a_0$, as shown in Figure 16 (b).

As follows from Figure 16 (b), for defect size $b = 0.2a_0$ there is only one minimum of the free energy, which smoothly shifts to higher values of lattice constant $a$ with increasing temperature. Therefore, there is no first-order phase transition in this case.

The main conclusion from Figures 16 (a) and (b) is that the first-order phase transition to the phase with large number of defects (condis phase) occurs at a certain temperature if the defect is geometrically large (and consequently the energy for the formation of such defect is also large).

This phase transition is accompanied by the discontinuous change of the lattice parameter $a$ and concentration of defects $n_d$, as shown in Figure 17. On the contrary, there is no first-order phase transition if the defect size $b$ is small and the lattice parameter and defect concentration increase smoothly.

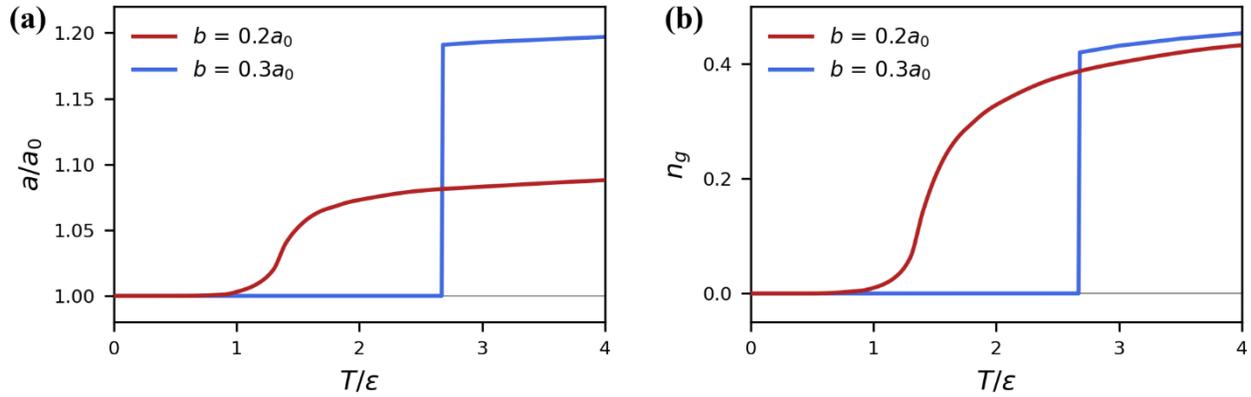

Figure 17. Dependencies of the lattice parameter $a$ (a) and fraction of defects (b) on temperature for defect sizes $b = 0.2a_0$ and $b = 0.3a_0$.

It is interesting to compare the results presented in Figure 12 with the results of the simplified model shown in Figure 17. Both the simplified model and more sophisticated analytical model (calibrated for the polyethylene) predict a qualitatively similar change in the behavior of $a(T)$ and fraction of defects $n_g(T)$, with a decrease of the defect size. This confirms that simplified model includes all the key physics that describes the phase transition to the condis phase.

Therefore, there is a critical defect size, $b_c$, which must be exceeded in order for a first-order phase transition to occur. Our analysis shows that the critical defect size and phase transition temperature are very sensitive to the parameters of the vdW interactions between the polymer chains and to the type of packaging of the chains in the crystal. Therefore, in order to obtain the correct phase transition temperature and the behavior of thermodynamic quantities for a particular polymer, it is necessary to take into account the complex details of interactions in the polymer that are omitted in the simplified model. In this case one should use the physics-based analytical model, developed in Section 3, the parameters of which have a clear physical meaning and are calibrated in accordance with the MD results.

In conclusion of this section, the simplified model indicates that a relatively large geometric size of defects in polymer chains (caused by gauche dihedrals) is a necessary condition for the first-order phase transition to the condis phase. When this condition is satisfied, the phase transition temperature can be approximately estimated as the potential energy of defect.

## 6. DISCUSSION AND CONCLUSIONS

A physics-based model of the phase transition of a polymer crystal from the crystalline to the condis phase is presented in the current paper. In the crystalline phase, all monomers of polymer chains are located in the nodes of the crystal lattice. In a condis crystal, despite the absence of a crystalline translational symmetry, the chains, irregular in the short range, are stretched in the long range along the axes, that are parallel to each other and form a dense hexagonal packing. Thus, in the case of polymer condis crystals, the positional and orientational orders are present only partially, in some specific sense. Nowadays many polymers in which the condis phase is observed are known. Among them there is a polyethylene, the simplest of known polymers. Therefore, we chose polyethylene as the object of modeling.

Based on the observations obtained from our MD simulations, the following physical picture of the phase transition to the condis phase has been established. The value of the lattice parameter, describing average distance between neighboring chains, increases with increasing temperature due to thermal expansion of the crystal. That results in an increase of the fraction of *gauche* dihedrals in the crystal, because the energy of *gauche* dihedral formation exponentially decreases with increase of the lattice parameter. Therefore, the thermal expansion of the crystal results in an increase of the conformational entropy of the polymer chains, which decreases the free energy of the crystal. On the other hand, the thermal expansion results in an increase of the vdW energy of inter-chain interaction (elastic energy), which increases the free energy of the crystal. Therefore, there is a competition between the conformational entropy of the polymer chains and the elastic energy of vdW inter-chain interaction at thermal expansion of the crystal. At low temperature, the equilibrium state of the polyethylene is the monoclinic crystalline phase with low fraction of the *gauche* dihedrals and relatively small value of the lattice parameter. At high temperature, the equilibrium state of the polyethylene is the condis phase with high fraction of the *gauche* dihedrals and large value of the lattice parameter. At the phase transition temperature, the free energies of these two states become equal to each other, and the crystal undergoes the first-order phase transition from the monoclinic to the condis phase under heating. At this phase transition, both the fraction of *gauche* dihedrals and the lattice parameter undergo a discontinuous change.

The total entropy change at the phase transition is approximately equal to the conformational entropy change of the polymer chains in the crystal, which was calculated from MD trajectories using the approximate analytical equation (2). This implies that the driving force of the phase transition from the monoclinic phase to the condis phase is an increase in the conformational entropy of polymer chains.

Based on the above physical picture, an analytical model of the phase transition was developed. The Gibbs free energy of a polymer crystal was constructed as a function of temperature and transverse lattice parameter. The value of the transverse lattice parameter for given temperature is calculated from the minimization of the Gibbs free energy. The analytical form of the Gibbs free energy includes elastic energy and the conformational entropy, which were extracted from the total energy and the total entropy calculated in MD simulations. The major contribution to the total energy in MD simulations is given by the internal degrees of freedom of a polymer chain consisting of rigid valence bonds, valence angles, and dihedral angles. Since these degrees of freedom are not involved in the phase transition, they were approximated as harmonic oscillators and subtracted from the total energy and total entropy.

The so obtained elastic energy and conformational entropy contain two contributions: a contribution that comes from the *trans* dihedrals only and an additional contribution from *gauche* dihedrals. To describe these contributions, a simple two-level model is used, in which it is assumed that the *trans* and *gauche* dihedrals are distributed along the chain randomly. The derived expression for the free energy includes the additional energy of *gauche* dihedrals that depends on the lattice parameter. The dependence of the energy of *gauche* dihedral on the lattice parameter was extracted form MD simulations. It turned out that the energy of *gauche* dihedrals looks like a smooth function if it is plotted vs. lattice parameter, and this function was approximated with a good accuracy by a decreasing exponential dependence.

The obtained analytical expression for the Gibbs free energy correctly describes the dependence of the transverse lattice parameter and the fraction of *gauche* dihedrals on temperature, because it is constructed using MD results. To verify the developed analytical model, the MD simulations of the polyethylene crystal were performed under applied external pressure 500 and 1000 atm. It was observed that the phase transition temperature increases with increasing pressure, as it should according to the Clapeyron-Clausius equation. The predictions of the analytical model for the pressure of 500 and 1000 atm are in perfect agreement with the results of MD simulations for the same pressure values.

In order to clearly and simply explain the physics of the phase transition to the condis phase, the analytical model has been significantly reduced, so that the resulting simplified model includes only some general properties of polymer chains playing a key role in the formation of condis phase. That is, the simplified model does not pretend to describe accurately any particular polymer. Instead, it is a minimal model that helps to understand the root cause of the first-order phase transition to the condis phase in the simplest possible way.

The simplified model describes a single polymer chain in an imaginary potential tube formed by van der Waals interactions with neighboring chains. The polymer chain contains a small fraction of randomly distributed "defects" that have a certain size, which is a parameter of the model. The energy required to form such a defect decreases rapidly with increasing distance between neighboring chains. This leads to a first-order phase transition at a certain temperature to the condis phase with a larger distance between neighboring chains and a higher fraction of defects. However, if the defect size is small, there is no first-order phase transition and the lattice parameter and defect fraction increase smoothly. Therefore, there is a critical defect size, which must be exceeded in order for a first-order phase transition to occur. When this condition is satisfied, the phase transition temperature can be approximately estimated as the potential energy of defect.

The proposed analytical approach can be extended to other polymers, where the condis phase exists. As an example of such polymers, vinylidene difluoride-based copolymers exhibiting a large electrocaloric effect may be mentioned. In addition, the proposed physics-based approach can be generalized to describe the phase transition of polymers into a melt.

ACKNOWLEDGMENT

The authors gratefully acknowledge the Joint Supercomputer Center of the Russian Academy of Sciences for computational resources granted. The authors are grateful to Dr. Sergei Burlatsky for useful discussions.

The present work was funded by the Ministry of Science and Higher Education, Russian Federation (Research theme state registration number 125020401357-4).